\documentstyle[aps,prl,floats,epsf]{revtex}

\def\Journal#1#2#3#4{{#1} {\bf #2}, #3 (#4)}

\def\NCA{Nuovo Cimento}

\def\NIMA{{Nucl. Instr. Meth.} A}

\def\PRL{Phys. Rev. Lett.}

\def\ADV{Adv. Space Res.}
\def\APJ{Astrophys. J.}
\def\APP{Astropart. Phys.}

\newcommand{\etal}{{\it et al.}}

\begin{document}

\draft

\wideabs{

\title{Measurements of Cosmic-ray Low-energy 
Antiproton and Proton Spectra in a Transient
Period of the Solar Field Reversal}
\author{
Y.\thinspace Asaoka$^{1}$,
Y.\thinspace Shikaze$^{1}$,
K.\thinspace Abe$^{1}$,
K.\thinspace Anraku$^{1}$,
M.\thinspace Fujikawa$^{1}$,
H.\thinspace Fuke$^{1}$,
S.\thinspace Haino$^{1}$,
M.\thinspace Imori$^{1}$,
K.\thinspace Izumi$^{1}$,
T.\thinspace Maeno$^{2}$,
Y.\thinspace Makida$^{3}$,
S.\thinspace Matsuda$^{1}$,
N.\thinspace Matsui$^{1}$,
T.\thinspace Matsukawa$^{2}$,
H.\thinspace Matsumoto$^{1}$,
H.\thinspace Matsunaga$^{1,}$\cite{Tsukuba},
J.\thinspace Mitchell$^{4}$,
T.\thinspace Mitsui$^{2,}$\cite{Tohoku},
A.\thinspace Moiseev$^{4}$,
M.\thinspace Motoki$^{1,}$\cite{Tohoku},
J.\thinspace Nishimura$^{1}$,
M.\thinspace Nozaki$^{2}$,
S.\thinspace Orito$^{1,}$\cite{Oriton},
J.\thinspace F.\thinspace Ormes$^{4}$,
T.\thinspace Saeki$^{1}$,
T.\thinspace Sanuki$^{1}$,
M.\thinspace Sasaki$^{3}$,
E.\thinspace S.\thinspace Seo$^{5}$,
T.\thinspace Sonoda$^{1}$,
R.\thinspace Streitmatter$^{4}$,
J.\thinspace Suzuki$^{3}$,
K.\thinspace Tanaka$^{3}$,
K.\thinspace Tanizaki$^{2}$,
I.\thinspace Ueda$^{1}$,
J.\thinspace Z.\thinspace Wang$^{5}$,
Y.\thinspace Yajima$^{6}$,
Y.\thinspace Yamagami$^{6}$,
A.\thinspace Yamamoto$^{3}$,
Y.\thinspace Yamamoto$^{1}$,
K.\thinspace Yamato$^{2}$,
T.\thinspace Yoshida$^{3}$,
and
K.\thinspace Yoshimura$^{3}$
}
\address{
$^{1}$The University of Tokyo, Tokyo, 113--0033 JAPAN\\
$^{2}$Kobe University, Kobe, Hyogo, 657--8501 JAPAN\\
$^{3}$High Energy Accelerator Research Organization (KEK),
Tsukuba, Ibaraki 305--0801, JAPAN\\
$^{4}$National Aeronautics and Space Administration, Goddard
Space Flight Center (NASA/GSFC), Greenbelt, MD 20771, USA\\
$^{5}$University of Maryland, College Park, MD 20742, USA\\
$^{6}$The Institute of Space and Astronautical Science (ISAS),
Sagamihara, Kanagawa 229--8510, JAPAN
}
\date{\today}
\maketitle

\begin{abstract}
The energy spectra of cosmic-ray low-energy antiprotons ($\bar{p}$'s)
and protons ($p$'s) have been measured by BESS in 1999 and 2000, during 
a period covering the solar magnetic field reversal.
Based on these measurements,
a sudden increase of the $\bar{p}/p$ flux ratio following the solar
magnetic field reversal was observed, 
and it generally agrees with a drift model of the solar modulation.
\end{abstract}

\pacs{PACS numbers: 98.70.Sa, 96.40.Kk, 95.85.Ry}

}
Much of the real underlying physics of the Sun is related to the 22 yr
solar magnetic cycle with recurrent positive and
negative phases.
The magnetic field polarity reverses when the solar activity is maximum,
and the global magnetic field profile reverses in the
heliosphere.
The most recent field reversal happened in the beginning of 2000.
The solar modulation of cosmic rays is caused by expanding solar wind,
which spreads out 
locally irregular magnetic field and therefore modifies energy spectra
of the cosmic rays entering the heliosphere.
The positive and negative particles drift in opposite directions,
taking different routes to arrive at the Earth in the heliospheric
magnetic field.
The charge-sign dependence is, therefore, a natural
consequence~\cite{JOK81}, 
on top of the common time dependent change in the overall modulation.
It explains 
alternate appearances of ''flat'' and ''peaked'' periods in neutron
monitor data around solar minima.
Recent works~\cite{Burger,BI99,Mosk01} also
indicated that the drift produces non-negligible effects
between the positive and negative particles even during
the high solar activity.
This view is supported by measurements of temporal variation
in cosmic-ray ratios~\cite{GA91_RA97}, such as electrons to helium
nuclei (He) and electrons to protons ($p$'s), where the 
largest variation is associated with the solar magnetic field reversal.
Among various cosmic-ray pairs, antiprotons ($\bar{p}$'s) and $p$'s
are ideal~\cite{BI99} for understanding drift effects under
the change in overall modulation level, because they are different
only in charge sign. 

In the last solar minimum period, the BESS experiment revealed that
the cosmic-ray $\bar{p}$ spectrum has a distinct peak around 2 GeV
\cite{OR00}, which is a characteristic feature 
of secondary $\bar{p}$'s produced by cosmic-ray interactions with
interstellar (IS) gas. It has become evident that
$\bar{p}$'s are predominantly secondary in origin, because
several recent calculations of the secondary spectrum basically agree
with observations in their absolute values and spectral shapes
\cite{BI99,Mosk01,Bergstrom99a,Donato01}. 

We report here new measurements of cosmic-ray $\bar{p}$ and $p$ fluxes 
and their ratios in the energy range from 0.18 to 4.2 GeV
collected in two BESS balloon flights carried out in
1999 and 2000, when the solar activity was maximum. 
Based on the solar magnetic field data~\cite{WSO}, the Sun's polarity 
reversed between these two flights~\cite{CL01}. 
With our full set of data~\cite{OR00,YO95,MAT98,MA01}, we discuss
the temporal variation of the $\bar{p}$ flux 
and $\bar{p}$/$p$ ratio covering the solar minimum, the maximum, and
the field reversal.

The BESS spectrometer was designed~\cite{OR87,YA88} and
developed~\cite{YA94,AJ00,AS98,SH00} as a high-resolution 
spectrometer.
A uniform field of 1 Tesla is produced by a thin superconducting solenoid
\cite{MA95}, and the field region is filled with tracking detectors.
This geometry results in an acceptance of 0.3 m$^2$sr.
Tracking is performed by fitting up to 28 hit-points in drift
chambers, resulting in a magnetic-rigidity ($R \equiv Pc/Ze$)
resolution of 0.5\% at 1 GV.
The upper and lower scintillator-hodoscopes~\cite{SH00} provide
time-of-flight and two d$E$/d$x$ measurements. 
Time resolution of each counter is 55 ps, resulting in a $1/\beta$
resolution of 0.014, where $\beta$ is defined as particle velocity
normalized by the speed of light.
The instrument also incorporates a threshold-type Cherenkov
counter~\cite{AS98} with a silica-aerogel radiator ($n$=1.02) that can 
reject $e^{-}$ and $\mu^{-}$ backgrounds by a factor of 4000 and
identify $\bar{p}$'s from such backgrounds up to 4.2 GeV.
In addition to biased trigger modes~\cite{MA01,AJ00} selecting
negatively-charged particles preferentially, one of every 60 (30)
first-level triggered events were recorded to provide unbiased samples
in 1999 (2000). Note that the first-level trigger was provided by
a coincidence between the top and bottom scintillators, with the
threshold set at 1/3 of the pulse height from minimum ionizing
particles. 

\begin{figure}[t]
\centerline{\epsfxsize=8.4cm \epsffile{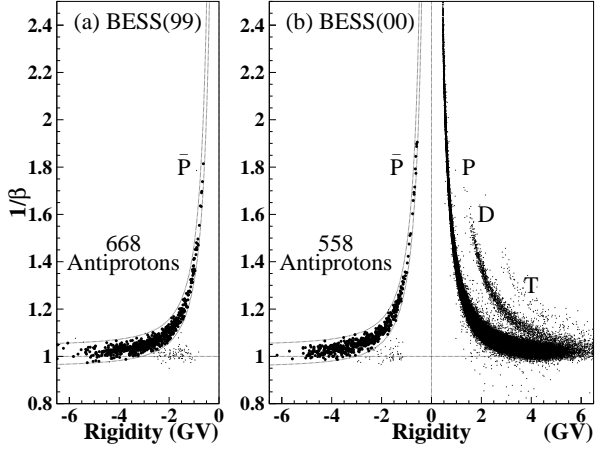}}
\caption{
The identification plots of $\bar{p}$ events for (a) 1999 and 
(b) 2000 flights. The dotted curves define the $\bar{p}$ mass bands.
}
\label{fig:id}
\end{figure}
The experiments were carried out in northern Canada, Lynn Lake to
Peace River, where the
geomagnetic cutoff rigidity ranges from 0.3 to 0.5 GV.
Data for flux measurements were taken
for live-time periods of 100,471 and 90,765 sec at altitudes above 34 km
(residual air of 4.3 and 5.0 g/cm$^2$ on average) in 1999 and 2000,
respectively. 
Data during ascent were also collected at various altitudes, in order
to estimate a background from atmospheric secondary $p$'s.

\widetext

\begin{table}[b]
\caption{
Antiproton fluxes 
(in $10^{-2}$ m$^{-2}$s$^{-1}$sr$^{-1}$GeV$^{-1}$), proton fluxes 
(in $10^{2}$ m$^{-2}$s$^{-1}$sr$^{-1}$GeV$^{-1}$), 
    and $\bar{p}/p$ ratio (in unit of $10^{-5}$)
at TOA.
$T$ (in GeV) defines the kinetic energy bins.
$N_{\bar{p}}$ is the number of observed antiprotons.
}
\label{tab:sum}
\renewcommand{\arraystretch}{1.245}
\begin{tabular}{c|crrr|crrr}

 & 
 & 
 & \multicolumn{1}{c}{BESS 1999}
 & 
 & 
 & 
 & \multicolumn{1}{c}{BESS 2000}
 & \\
$T$ (GeV)
    & \multicolumn{1}{c}{ $N_{\bar{p}}$}
& \multicolumn{1}{c}{ $\bar{p}$ flux \qquad}
& \multicolumn{1}{c}{ $p$ flux \qquad }
& \multicolumn{1}{c|}{ $\bar{p}/p$ ratio \qquad }
& \multicolumn{1}{c}{ $N_{\bar{p}}$ }
& \multicolumn{1}{c}{ $\bar{p}$ flux \qquad}
& \multicolumn{1}{c}{ $p$ flux \qquad}
& \multicolumn{1}{c}{ $\bar{p}/p$ ratio \qquad} \\
\tableline
0.18 - 0.28
 &     2
 & $0.22^{+0.28\,+0.02}_{-0.15\,-0.02} \quad $
 & $10.42^{+0.09\,+1.01}_{-0.09\,-1.01} \quad $
 & $0.21^{+0.26\,+0.02}_{-0.14\,-0.02} \quad $
 &     7
 & $0.75^{+0.41\,+0.08}_{-0.34\,-0.08} \quad $
 & $1.71^{+0.04\,+0.35}_{-0.04\,-0.35} \quad $
 & $4.38^{+2.37\,+0.94}_{-1.97\,-0.94} $ \\
0.28 - 0.40
 &     9
 & $0.57^{+0.28\,+0.04}_{-0.20\,-0.04} \quad $
 & $11.79^{+0.07\,+0.60}_{-0.07\,-0.60} \quad $
 & $0.49^{+0.24\,+0.04}_{-0.17\,-0.04} \quad $
 &     6
 & $0.40^{+0.26\,+0.03}_{-0.18\,-0.03} \quad $
 & $2.28^{+0.03\,+0.22}_{-0.03\,-0.22} \quad $
 & $1.76^{+1.16\,+0.21}_{-0.77\,-0.21} $ \\
0.40 - 0.56
 &    28
 & $1.25^{+0.31\,+0.07}_{-0.27\,-0.07} \quad $
 & $11.76^{+0.06\,+0.44}_{-0.06\,-0.44} \quad $
 & $1.07^{+0.26\,+0.06}_{-0.23\,-0.06} \quad $
 &     8
 & $0.29^{+0.19\,+0.04}_{-0.16\,-0.04} \quad $
 & $2.62^{+0.02\,+0.14}_{-0.02\,-0.14} \quad $
 & $1.12^{+0.74\,+0.17}_{-0.60\,-0.17} $ \\
0.56 - 0.78
 &    27
 & $0.71^{+0.22\,+0.07}_{-0.20\,-0.07} \quad $
 & $10.51^{+0.05\,+0.31}_{-0.05\,-0.31} \quad $
 & $0.67^{+0.21\,+0.06}_{-0.19\,-0.06} \quad $
 &    36
 & $1.09^{+0.27\,+0.08}_{-0.23\,-0.08} \quad $
 & $2.74^{+0.02\,+0.10}_{-0.02\,-0.10} \quad $
 & $4.00^{+0.99\,+0.29}_{-0.83\,-0.29} $ \\
0.78 - 1.10
 &    63
 & $1.20^{+0.21\,+0.07}_{-0.20\,-0.07} \quad $
 & $8.76^{+0.03\,+0.25}_{-0.03\,-0.25} \quad $
 & $1.37^{+0.24\,+0.08}_{-0.22\,-0.08} \quad $
 &    67
 & $1.35^{+0.24\,+0.08}_{-0.22\,-0.08} \quad $
 & $2.64^{+0.01\,+0.08}_{-0.01\,-0.08} \quad $
 & $5.11^{+0.90\,+0.30}_{-0.84\,-0.30} $ \\
1.10 - 1.53
 &   112
 & $1.85^{+0.24\,+0.11}_{-0.23\,-0.11} \quad $
 & $6.70^{+0.03\,+0.20}_{-0.03\,-0.20} \quad $
 & $2.77^{+0.36\,+0.16}_{-0.34\,-0.16} \quad $
 &    77
 & $1.20^{+0.21\,+0.08}_{-0.19\,-0.08} \quad $
 & $2.37^{+0.01\,+0.07}_{-0.01\,-0.07} \quad $
 & $5.05^{+0.88\,+0.32}_{-0.82\,-0.32} $ \\
1.53 - 2.15
 &   138
 & $1.89^{+0.22\,+0.11}_{-0.21\,-0.11} \quad $
 & $4.60^{+0.02\,+0.16}_{-0.02\,-0.16} \quad $
 & $4.12^{+0.48\,+0.23}_{-0.46\,-0.23} \quad $
 &   102
 & $1.33^{+0.20\,+0.08}_{-0.19\,-0.08} \quad $
 & $1.92^{+0.01\,+0.08}_{-0.01\,-0.08} \quad $
 & $6.95^{+1.07\,+0.44}_{-1.00\,-0.44} $ \\
2.15 - 3.00
 &   162
 & $1.70^{+0.18\,+0.13}_{-0.17\,-0.13} \quad $
 & $2.90^{+0.01\,+0.11}_{-0.01\,-0.11} \quad $
 & $5.86^{+0.63\,+0.42}_{-0.60\,-0.42} \quad $
 &   131
 & $1.32^{+0.17\,+0.11}_{-0.16\,-0.11} \quad $
 & $1.44^{+0.01\,+0.05}_{-0.01\,-0.05} \quad $
 & $9.20^{+1.18\,+0.78}_{-1.12\,-0.78} $ \\
3.00 - 4.20
 &   127
 & $1.20^{+0.15\,+0.18}_{-0.14\,-0.18} \quad $
 & $1.70^{+0.01\,+0.06}_{-0.01\,-0.06} \quad $
 & $7.05^{+0.90\,+0.98}_{-0.85\,-0.98} \quad $
 &   124
 & $1.19^{+0.16\,+0.19}_{-0.15\,-0.19} \quad $
 & $1.00^{+0.01\,+0.04}_{-0.01\,-0.04} \quad $
 & $11.93^{+1.59\,+1.79}_{-1.50\,-1.79} $ \\
\end{tabular}
\end{table}

\renewcommand{\arraystretch}{1}
\narrowtext
Analysis was performed in the same way as described in Ref.~\cite{MA01}.
We applied the same selection criteria for $\bar{p}$'s and $p$'s
because non-interacting $\bar{p}$'s behave like $p$'s except for
deflection in the symmetrical configuration of BESS.
Figure~\ref{fig:id} shows $\beta^{-1}$ versus $R$ plots
for surviving events (for 1999, only the negative rigidity side is 
shown).
We see clean narrow bands of 668 and 558 $\bar{p}$ candidates at the
exact mirror position of the $p$'s for the 1999 and 2000 data,
respectively. 
The $\bar{p}$ bands are slightly contaminated by $e^{-}$ and $\mu^{-}$
backgrounds due to inefficiency of the aerogel Cherenkov counter, 
while contamination on the $p$ band was negligible.
We estimated the fraction
of $e^{-}$ and $\mu^{-}$ backgrounds to be $0$ \%($0$ \%), $0.6$ \%
($0.5$ \%), and $2.3$ \% ($1.5$ \%), respectively, at 0.3, 2, and 4 GeV
for 1999 (2000) data. 
Other backgrounds such as albedo, mis-measured positive-rigidity
particles, and ``re-entrant albedo'' were found to be negligible.

\widetext
\begin{figure*}[t]
  \vspace*{-0.9cm}
  \begin{center}
     \centerline{\epsfxsize=16cm \epsffile{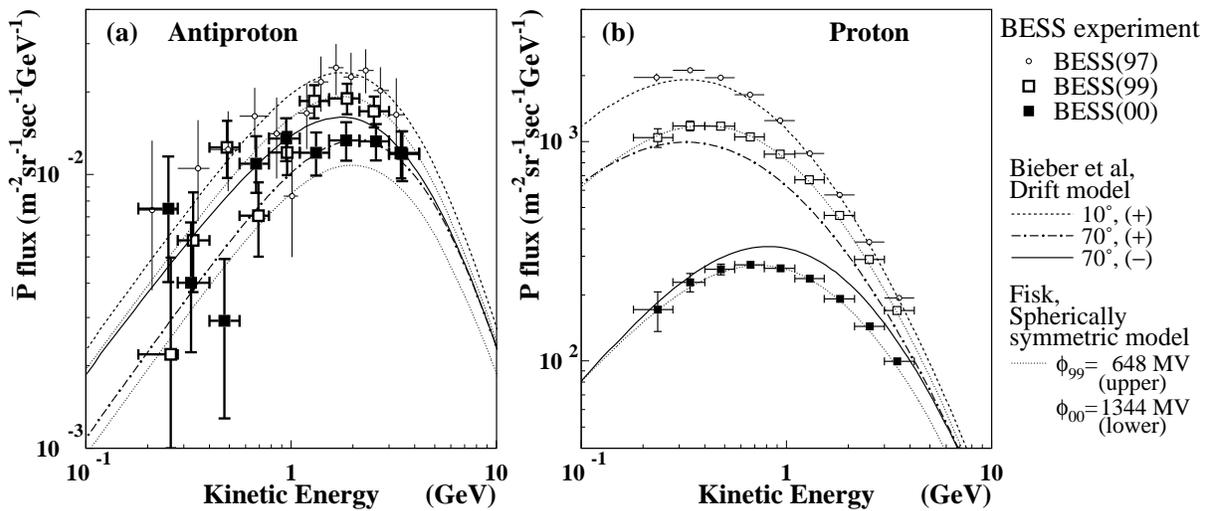}}
    \caption{
     The (a) $\bar{p}$ and (b) $p$ fluxes at TOA
     measured by BESS in 1999 and 2000 together with the previous data
     in 1997.
     The curves represent the calculations by the drift model
     \protect\cite{BI99} at a period of 10$^\circ$($+$), 
     70$^\circ$($+$), and 70$^\circ$($-$),
     which respectively represent solar minimum at
     positive phase (dashed line), solar maximum at positive phase
     (dash-dotted line), and solar maximum at negative phase (solid
     line).
     The dotted curves represent the calculations by the spherically
     symmetric model \protect\cite{FISK}.
	}
    \vspace*{-1.0cm}
    \label{fig:flx}
  \end{center}
\end{figure*}
\narrowtext

The survival probabilities for $\bar{p}$'s and $p$'s to traverse
residual air and the instrument without interactions and to pass
through all the 
selection criteria described above were estimated by a Monte Carlo (MC)
simulation. The MC code was tuned and verified by comparing the
simulation with an accelerator beam test of
the BESS detector~\cite{AS01}. 
The systematic uncertainty of the $\bar{p}$ interaction losses were
reduced to 5 \%. 
Estimates of the background from atmospheric secondary $\bar{p}$'s were
calculated~\cite{MI96,ST97,PF96} with the $p$ and
He fluxes measured by BESS as input. The subtraction amounts to 15
$\pm$ 5 \%  (17$\pm$6 \%), 23 $\pm$ 2 \% (30 $\pm$ 2 \%), and 26 $\pm$
8 \% (28 $\pm$ 9 \%) for 0.3, 2, and 4 GeV in 1999  (2000),
respectively, where the errors correspond to the maximum difference
among the calculations.
The background from atmospheric secondary $p$'s was estimated according
to the calculation~\cite{PA96}.
The parameters used in the calculation were tuned and verified
\cite{SH01} by comparing the calculated fluxes at various air depths
with the atmospheric growth curve of $p$'s measured by BESS. 
The subtraction amounts to 27 $\pm$ 4 \% 
(55 $\pm$ 8 \%), 10 $\pm$ 2 \% (22 $\pm$ 3 \%), and 4 $\pm$ 1 \% (8
$\pm$ 1 \%) for 0.2, 0.5, and 1 GeV in 1999  (2000), respectively,
where the errors correspond to the maximum
difference between observed and calculated atmospheric
growth curve of the $p$'s. 
Table~\ref{tab:sum} gives the resultant
fluxes of $\bar{p}$'s and $p$'s 
and their flux ratios, $\bar{p}/p$, at the top of
the atmosphere (TOA) with the statistical (first) and systematic 
(second) errors.

Figure~\ref{fig:flx} shows the $\bar{p}$ and $p$ fluxes 
in 1999 and 2000, together with the data~\cite{OR00} taken in 1997 at
the solar minimum period. 
The error bars represent quadratic sums of
\mbox{ } \hspace*{8.0cm} \vspace{6.5cm} \mbox{ }
the statistical and systematic errors, as well as in the 
following figures. 
The dotted curves represent the modulated spectra according to the
standard 
spherically symmetric approach~\cite{FISK}, 
in which the modulation is
characterized by a single parameter ($\phi$) irrespective of the Sun's
polarity. In each year, $\phi$ was determined to fit the $p$ spectrum
measured by BESS assuming the IS spectrum in Ref.~\cite{BI99}.
The same $\phi$ was applied to modulate the $\bar{p}$ flux.
In both plots, dashed, dash-dotted, and solid curves
represent modulated spectra according to a steady-state drift
model~\cite{BI99} in which the modulation is characterized by a tilt
angle ($\alpha$) of the heliospheric current sheet and the Sun's
magnetic polarity (denoted as $+$/$-$). Note that periods at 
10$^\circ$($+$), 70$^\circ$($+$), and 70$^\circ$($-$) roughly 
correspond to the years of 1997, 1999, and 2000, respectively.
However, it is shown in Fig.~\ref{fig:flx} that the actual level of
overall modulation at the date of observation was lower
(corresponding to a smaller value of $\alpha$) in 1999
and higher (larger $\alpha$) in 2000. 
Thus, the predicted effect of the drifts in Fig.~\ref{fig:flx}(a)
might be hidden under the change in overall modulation level.
The drift model needs to be fine-tuned to reproduce the $\bar{p}$ and
$p$ spectra. 
The retardation of the modulation effect due to the gradually
expanding heliospheric current sheet~\cite{RO90} might also be
concerned.

\begin{figure}[b]
\vspace*{-.6cm}
\centerline{\epsfxsize=8.0cm \epsffile{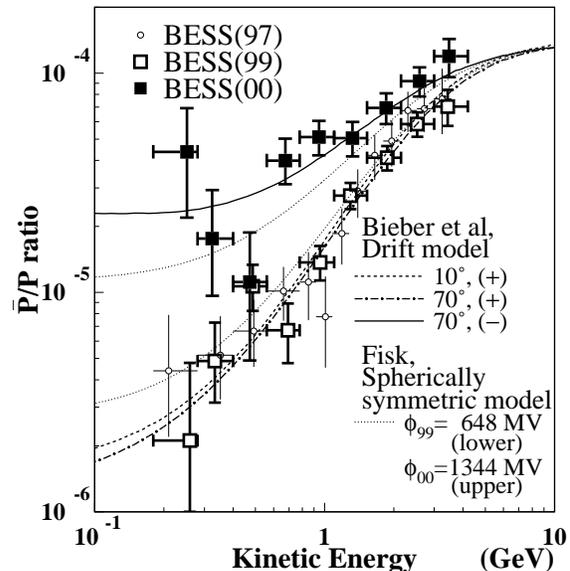}}
\caption{
The $\bar{p}/p$ flux ratios measured by BESS in 1999 and 2000 with the
previous data in 1997.
The curves are calculations of the ratio at various solar activity
\protect\cite{BI99,FISK}.
}
\label{fig:pbp}
\end{figure}
Figure~\ref{fig:pbp} shows the $\bar{p}/p$ ratios in 1999 and 2000 in
which the above mentioned difficulty is partly canceled.
In general, $\bar{p}$'s suffer less modulation than $p$'s because of
their relatively hard IS spectrum.
Thus, the spherically symmetric model predicts an increase in $\bar{p}/p$
towards the solar maximum period~\cite{LA97}, mainly due to a decrease
of the $p$ flux.  
The drift model predicts a much larger increase in $\bar{p}/p$
reflecting the charge-sign dependence.
It should be noted that a recent independent work~\cite{Mosk01}
has shown qualitatively the same feature in $\bar{p}/p$ while these two
drift model calculations were different in IS spectra and some modulation
parameters.
As shown in Fig.~\ref{fig:pbp}, the measured $\bar{p}/p$ ratios are in 
better agreement with the drift model than the spherically symmetric model.

\begin{figure}[t]
\vspace*{-0.6cm}
\centerline{\epsfxsize=8cm \epsffile{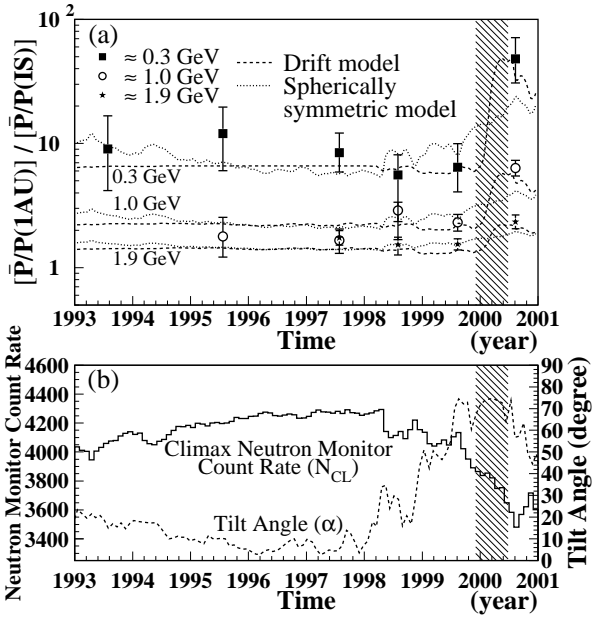}}
\caption{(a) Temporal variation of the $\bar{p}/p$ ratios (relative to
the ratio in IS) at 0.3 (closed squares), 1.0 (open circles), and 1.9
GeV (closed stars) measured by BESS since 1993. 
Curves represent the calculated temporal variation of the
ratios (see text).
(b) Temporal variation of $N_{\rm CL}$ and $\alpha$.
In both plots, shaded area indicates the period of the solar field
reversal. 
}
\label{fig:mod}
\end{figure}
Combining all the previous BESS measurements since 1993
\cite{OR00,YO95,MAT98,MA01}, 
annual variations of the $\bar{p}/p$
ratios (relative to the ratio in IS) at energies of 0.2 -- 0.4 GeV
(closed square, denoted as 
0.3 GeV), 0.8 -- 1.3 GeV (open circle, 1.0 GeV), and 1.5 -- 2.5 GeV
(closed star, 1.9 
GeV) are shown in Fig.~\ref{fig:mod}(a). These data show no distinctive
year-to-year variation before the solar field reversal indicated by
the shaded area in Fig~\ref{fig:mod}(a).
The curves represent calculations of the $\bar{p}/p$ time variation 
according to the drift model (dashed) and the spherically symmetric
model (dotted) for 0.3, 1.0, and 1.9 GeV from top to bottom.
We obtained those $\phi$'s at the dates of the BESS flights ($\phi_{\rm
BESS}$) and then we evaluated $\phi$'s at any moments using a linear
relation between $\phi_{\rm BESS}$ and $N_{\rm CL}$, where $N_{\rm CL}$
is the monthly averaged count rates of the Climax Neutron Monitor
\cite{NEUT}. 
The historical trend of $N_{\rm CL}$ is indicated by the solid
histogram in Fig.~\ref{fig:mod}(b). 
In Ref.~\cite{BI99}, the $\bar{p}/p$ ratios are given as a function of
the tilt angle, $\alpha$, rather than the actual time.
In order to draw the dashed curve in Fig.~\ref{fig:mod}(a), we take
$\alpha$ as the mean position of the maximum latitudinal extent of the
current sheet~\cite{TILT} taking no account of the retardation of
the modulation effect.
Note that $\alpha$ is allowed to reach 90$^\circ$ 
in the shaded area to avoid discontinuity in $\bar{p}/p$ at the field
reversal. 
The drift model predicts that the $\bar{p}/p$ ratio is stable during the
positive phase and shows a sudden increase after the
positive-to-negative solar field reversal.
Our measurements generally agree with the drift model.

As a conclusion, we have measured the temporal variation of
the $\bar{p}$ and $p$ fluxes covering the solar minimum, the maximum,
and the solar magnetic field reversal. 
We observed stable $\bar{p}/p$ ratios in the positive
polarity phase through 1999 and a sudden increase in 2000 following the
solar field reversal.
The continuous observation has enabled a crucial test of drift effects
in the solar modulation including a solar maximum period.
It will motivate further development of models with more realistic
parameters~\cite{Mosk01} and time dependence~\cite{BU01}.
Furthermore, detailed understanding of solar modulation is
inevitably important for investigating the origin of
$\bar{p}$'s~\cite{YA00}, 
especially in the very low energy region below 0.5 GeV, where 
slightly excessive $\bar{p}$ fluxes relative to the theoretical
calculations were observed during the solar minimum~\cite{OR00,MAT98}.

We thank NASA and NSBF for the balloon expedition,
and KEK and ISAS for continuous support.
This work was supported by Grant-in-Aid for Scientific
Research, MEXT in Japan;
and by NASA in the USA.
Analysis was performed using the computing facilities at
ICEPP, the University of Tokyo.

\end{document}